\documentstyle[11pt,epsfig]{article}

\def\laq{~\raise 0.4ex\hbox{$<$}\kern -0.8em\lower 0.62
ex\hbox{$\sim$}~}
\def\gaq{~\raise 0.4ex\hbox{$>$}\kern -0.7em\lower 0.62
ex\hbox{$\sim$}~}

\def\beq{\begin{equation}}
\def\eeq{\end{equation}}
\def\bea{\begin{eqnarray}}
\def\eea{\end{eqnarray}}
\def\bean{\begin{eqnarray*}}
\def\eean{\end{eqnarray*}}

\def \pa {\partial}

\def \la {\lambda}
\def \La {\Lambda}
\def \Da {\Delta}
\def \b {\beta}

\def \da {\delta}
\def \ep {\epsilon}
\def \r {\rho}

\def \Om {\Omega}
\def \noi {\noindent}

\def \ep {\epsilon}
\def \Mp {M_{\rm P}}

\begin{document}

\begin{titlepage}

\begin{flushright}
BA-TH/04-491\\
 Bicocca-FT-04-9\\
astro-ph/0407573
\end{flushright}

\vspace*{0.6 cm}

\begin{center}

\Huge
{Fitting Type Ia supernovae \\ 
with coupled dark energy}

\vspace{1cm}

\large{L. Amendola$^1$, M. Gasperini$^{2,3}$ 
and F. Piazza$^4$}

\normalsize
\vspace{.2in}

{\sl $^1$INAF/Osservatorio Astronomico di Roma, \\
 Via Frascati 33, 00040 Monteporzio Catone (Roma), Italy}

{\sl $^2$Dipartimento di Fisica, Universit\`a di Bari, \\ 
Via G. Amendola 173, 70126 Bari, Italy}

{\sl $^3$Istituto Nazionale di Fisica Nucleare, Sezione di Bari, Italy}

{\sl $^4$Dipartimento di Fisica and INFN, Universit\`{a} di Milano
Bicocca, \\
 Piazza delle Scienze 3, I-20126 Milano, Italy}

\vspace*{1.3cm}

\begin{abstract}
We discuss the possible consistency of the recently discovered Type Ia supernovae at $z>1$ with models in which dark energy is strongly coupled to a significant fraction of dark matter, and in which an (asymptotic) accelerated phase exists where dark matter and dark energy scale in the same way. Such a coupling has been suggested for a possible solution of the coincidence problem, and is also motivated by string cosmology models of ``late time" dilaton interactions.  Our analysis shows that, for coupled dark energy models, the recent data are still consistent with acceleration starting as early as at $z=3$ (to within 90\% c.l.), although at the price of a large ``non-universality" of the dark energy coupling to different matter fields. Also, as opposed to uncoupled models which seem to prefer a ``phantom'' dark energy, we find that a large amount of coupled dark matter is compatible with present data only if the dark energy field has a conventional equation of state $w>-1$. 
\end{abstract}
\end{center}

\end{titlepage}

\newpage

There is increasing evidence that our Universe is presently in a state
of cosmic acceleration \cite{1} (or ``late-time\char`\"{}
inflation), i.e. that its energy is dominated by a component with
negative (enough) pressure, dubbed {}``quintessence\char`\"{}, or
dark energy. The energy density of such a cosmic component is, at
present, comparable with that of the more conventional (pressureless) dark matter component, although their evolution equations can be widely different, in principle. 

In order to alleviate such a {}``coincidence\char`\"{} problem \cite{2}, without fine-tuning and/or \emph{ad hoc} assumptions
on the viscosity of the dark matter fluid \cite{3}, a direct (and
strong enough) coupling between dark matter (or at least a significant part of it) and dark energy has been proposed \cite{4}. Phenomenological models of this type have been widely studied \cite{4,5,5a,5b}, and shown to be compatible with various constraints following from supernovae observations, cosmological perturbation theory and structure formation.  They are also theoretically motivated by the identification of the dark energy field with the string-theory dilaton \cite{6,PT}, and by the assumption that dilaton loop corrections saturate in the strong coupling regime \cite{7}. 

A key feature of the coupled dark energy scenario is that it may allow a late-time attractor describing an accelerated phase, characterized by a frozen ratio $\rho _{\rm{dark~matter}}/\rho _{\rm{dark~energy}}$. This occurs, in models where a rolling scalar field $\phi$ plays the role of the cosmic dark energy, when the dark energy has an exponential potential and is exponentially coupled to scalar dark-matter fields (such an attractor property is lost, however, in the case of Yukawa-type interactions to fermionic dark matter \cite{10a}). For a successful scenario it is also required that $a)$ the present coupling of the scalar
to some exotic dark matter component is strong enough and approximately
constant, and $b)$ the present coupling to ordinary (baryonic) matter is \emph{weak} enough, in order to avoid testable violations of the equivalence principle. 

In addition, the cosmic dark energy field has to be either massless or ultra-light, with a mass
$m\sim H_{0}$, where $H_{0}\simeq 10^{-33}$ eV is the present  Hubble scale. For a scalar field gravitationally coupled to particles of mass $M$, on the other hand, there are quantum (radiative) corrections to the mass  \cite{8aa} of order $m \sim qM(\La /\Mp )$, where $q$ is the dimensionless coupling strength
and $\La /\Mp $ is the cut-off scale (typically, $1$ TeV), in Planck
units. This well known perturbative argument seems to suggest that both the coupling $q$ to ordinary baryonic matter (for which $M\sim 1$ Gev), and the value of $M$ for the coupled, exotic dark matter components (for which the models assumes $q\simeq 1$), 
need to be fine-tuned at extremely low values \cite{8a}. It should be stressed, however, that scalars with masses comparable
to the (four dimensional) curvature of the Universe could be safe from radiative corrections, thanks to a recently proposed mechanism based on the   AdS/CFT correspondence \cite{GP}. 

Quite independently from theoretical motivations, and from possible
interpretations in a string cosmology context, in this work we consider the basic scenario \cite{4} in which dark energy is parametrized by a
(canonically normalized) scalar field $\phi $, with exponential self-interaction
energy of slope $\mu \sqrt{2/3}$, 
\begin{equation}
V(\phi )=V_{0}e^{-\mu \sqrt{\frac{2}{3}}{\frac{\phi }{\Mp }}}.
\label{1}
\end{equation}
 Here $\mu >0$ and $\Mp =(8\pi G)^{-1/2}$ is the reduced Planck mass.  Neglecting radiation, as we are concerned
with the late evolutionary phase of our Universe, we can describe
the remaining gravitational sources as a cosmic fluid of non-relativistic,
{}``dust\char`\"{} particles, distinguishing however a {}``coupled\char`\"{}
and {}``uncoupled\char`\"{} component, with energy densities $\r {_{c}}$
and $\r {_{u}}$, respectively. Uncoupled matter (for instance, baryons)
satisfies the usual conservation equation \begin{equation}
\dot{\r {_{u}}}+3H\r {_{u}}=0,\label{2}\end{equation}
 while the coupled evolution of $\r {_{c}}$ and $\phi $ is described
by the system of equations \begin{eqnarray}
 &  & \dot{\r {_{c}}}+3H\r {_{c}}-\b {\sqrt{\frac{2}{3}}}{\frac{\dot{\phi }}{\Mp }}\r {_{c}}=0,\\
 &  & \ddot{\phi }+3H\dot{\phi }+{\frac{\pa V}{\pa \phi }}+\b {\sqrt{\frac{2}{3}}}{\frac{1}{\Mp }}\r {_{c}}=0,\label{4}
\end{eqnarray}
 with $\b {>}0$. The total energy density is finally normalized according
to the Einstein equation \begin{equation}
3\Mp ^{2}H^{2}=\rho _{\phi }+\rho _{c}+\rho _{u},\label{5}\end{equation}
 where $\r {_{\phi }}={\dot{\phi }^{2}/2}+V(\phi ).$
To make contact with previous work, we have used the notations of
\cite{4,5}, but we stress that the above equations also describe
the asymptotic {}``freezing\char`\"{} phase of a {}``running dilaton\char`\"{}
model \cite{6}, provided we identify the slope $\mu \sqrt{2/3}$
and the coupling parameter $\b {\sqrt{2/3}}$ of the present model
with the slope $\la $ and the dilaton charge $q$ of \cite{6} as
follows: \begin{equation}
\mu \sqrt{\frac{2}{3}}=\la , ~~~~~~~
\b {\sqrt{\frac{2}{3}}}={\frac{q\la }{2}}.\label{7}
\end{equation}

The above equations can be derived (for a conformally flat metric)
from the (Einstein-frame) action \begin{equation}
{S}\, \, =\, \, \int d^{4}x\sqrt{-g}\left[-\frac{M_P^{2}}{2}\, R+\frac{1}{2}(\nabla \phi )^{2}-V(\phi )\right]+{\mathcal{S}}_{c}[\phi ,\psi _{c},g_{\mu \nu }]\, +{\mathcal{S}}_{u}[\psi _{u},g_{\mu \nu }],\label{8}\end{equation}
 where $\psi _{c}$ and $\psi _{u}$ are the coupled and uncoupled
matter fields, respectively. The parameter $\b {}$ is thus related
to the (homogeneous) scalar charge density, $\da S_{c}/\da \phi $,
of coupled matter, as \begin{equation}
\b {\sqrt{\frac{2}{3}}}=-{\frac{\Mp }{\r {_{c}}\sqrt{-g}}}{\frac{\da S_{c}}{\da \phi }}.\label{9}\end{equation}
As shown in \cite{4,6}, such a system of equations is characterized
by an asymptotic accelerated regime in which the uncoupled component
has redshifted away, $\r {_{u}}= 0$, and all other terms of
the Einstein equations have the same scaling behaviour: \begin{equation}
V\propto \dot{\phi }^{2}\propto \r {_{\phi }}\propto \r {_{c}}\propto H^{2}\propto a^{-3(1+w)},
~~~~\qquad w=-\frac{\beta }{\beta +\mu } 
\label{10}
\end{equation}
(a wider
class of scalar field Lagrangians leading to similar attractor configurations
has been studied in \cite{PT}). Within this regime $\r {_{\phi }}/\r {_{c}}=$ const, and $\Om _{\phi }=\r {_{\phi }}/3\Mp ^{2}H^{2}$,
$\Om _{c}=\r {_{c}}/3\Mp ^{2}H^{2}$, are also constant. The evolution
is accelerated, with constant acceleration parameter, \begin{equation}
{\frac{\ddot{a}}{aH^{2}}}=1+{\frac{\dot{H}}{H}}={\frac{q-1}{q+2}}={\frac{2\b {-}\mu }{2\b {+}\mu }},\label{11}\end{equation}
 provided $2\b {>}\mu $ (or $q>1$ in the notations of \cite{6}).
The luminosity-distance relation, for this regime, is obtained from
the Hubble function \begin{equation}
H(z)=H_{0}(1+z)^{3(1+w)/2}\label{12}\end{equation}
 (see Eq. (\ref{10})), where $z(t)=a_{0}/a(t)-1$ is the red-shift
parameter. The corresponding Hubble diagram is thus completely controlled
by the dark energy parameter $w$, and its compatibility with previously
known Type Ia supernovae (SNe Ia) has been studied in \cite{5}. 

Motivated by the recent important discovery of many new SNe Ia at
$z>1$ \cite{9}, the aim of this paper is to re-discuss a possible
fitting of present supernovae data in the context of the above model
of coupled dark energy. Instead of assuming the asymptotic regime
as an appropriate description of our present cosmological state, here
we consider a {}``quasi-asymptotic\char`\"{} configuration, describing
a cosmological state in the vicinity of the attractor, and for which
the uncoupled component $\r {_{u}}$ is small, but nonzero. The function
$H(z)$, and the associated luminosity-distance relation, will then
be obtained by perturbing the {}``freezing\char`\"{} \cite{6} (or
{}``stationary\char`\"{} \cite{4}) configuration (\ref{10}), to
the linear order. Such a perturbed configuration will contain the
(critical) fraction of uncoupled matter, $\Om _{u}$, as a parameter
measuring the typical (phase-space) {}``distance\char`\"{} from the
asymptotic attractor with $\Om _{u}=0$ (a different model with coupled and uncoupled fraction of dark matter components has been recently considered also in \cite{5b}). 

The uncoupled matter density scales in time faster than $\r {_{c}}$,
namely $\r {_{u}}\sim a^{-3}\sim (1+z)^{3}$, exactly like the standard
dark matter component. Unlike in conventional models of uncoupled
quintessence, however, the fraction $\Om _{u}$ of uncoupled matter
is not to be identified with the total present fraction of non-relativistic
(baryon+dark matter) components, and is not necessarily close to $0.3$.
In the case discussed here, the present value of $\Om _{u}$ may range
from the minimum $\Om _{u}=\Om _{\rm{baryons}}\simeq 0.05$ (we
will refer to this as the {}``minimal model''), to a maximum amount
consistent with the clustered fraction of dust energy density $\Omega _{m}$.
As a reference value, we will assume $\Omega _{m}=0.3$. 

Such a different interpretation of $\Om _{u}$, as we shall see, has
two important consequences. The first one, already stressed in \cite{5},
is that the accelerated regime may have a longer past extension, starting
at early epochs characterized by $z>1$, which are instead forbidden
for the standard (i.e. uncoupled) quintessential models \cite{9}. 

The second one, which probably represents the main result of this
paper, is that a small enough value of $\Om _{u}$ is consistent with
present data only for $w>-1$. If we take, for instance, $\Om _{u}<0.2$,
we obtain $w>-1$ at the $99\%$ confidence level. More conventional
analyses, assuming a fraction of uncoupled dark matter near to or
larger than $\Om _{m}\simeq 0.3$, seem to favour instead $w<-1$
\cite{10}, although $w>-1$ is certainly still consistent with data.
Should future data  confirm these results, we would be left with the
choice between {}``phantom\char`\"{} models of dark energy \cite{11},
uncoupled to dark matter but plagued by severe quantum instabilities 
\cite{Carroll} (see, however, \cite{onemli}),
and dilaton-like models of dark energy \cite{4}-\cite{PT}, non-minimally
coupled to a significant fraction of dark matter. 

In order to perturb the dynamical system of coupled equations (\ref{2})-(\ref{5})
it proves convenient to introduce the dimensionless variables \begin{equation}
x^{2}=\frac{\dot{\phi }^{2}}{6\Mp ^{2}H^{2}}\, ,\qquad y^{2}=\frac{V}{3\Mp ^{2}H^{2}}\, ,\qquad u^{2}=\frac{\rho _{u}}{3\Mp ^{2}H^{2}}\, .\end{equation}
 Obviusly, $\Om _{c}=1-x^{2}-y^{2}-u^{2}$. Denoting with a prime
the derivative with respect to the evolution parameter $N=\ln (a/a_{0})$,
the coupled system of equations, for the model presented in the previous
section, can then be rewritten as \cite{4,12}: \begin{eqnarray}
x^{\prime } & = & -\frac{3}{2}\left(1-x^{2}+y^{2}\right)x-\mu y^{2}+\beta (1-x^{2}-y^{2}-u^{2}),\nonumber \\
y^{\prime } & = & \mu xy+\frac{3}{2}y\left(1+x^{2}-y^{2}\right),\nonumber \\
u^{\prime } & = & \frac{3}{2}u\left(x^{2}-y^{2}\right).\label{sys2}
\end{eqnarray}
 Since there is complete symmetry with respect to simultaneous sign
inversion of $\beta ,\mu $ and $\phi $, we will consider the case
$\mu >0$ only. It is also useful to write the equation for the derivative
of $H$: one has, from Eq. (\ref{5}), \begin{equation}
\frac{H'}{H}=-\frac{1}{2}(3+3x^{2}-3y^{2}).\label{hprime}\end{equation}

By equating the right hand sides of (\ref{sys2}) to zero one obtains
a system of algebraic equations whose solutions are characterized
by constant fractional densities. Of particular interest is the freezing
(or stationary) configuration, given by \begin{equation}
x_{0}=-\frac{3}{2(\beta +\mu )}\, ,\qquad y_{0}=\frac{\sqrt{9+4\beta \mu +4\beta ^{2}}}{2(\beta +\mu )}\, ,\qquad u_{0}=0.\label{att}\end{equation}
 This point exists for $\beta >0$ and $\mu >(-\beta +\sqrt{18+\beta ^{2}})/2$,
is accelerated for $\mu <2\beta $, and is a global attractor \cite{4}.
On this solution, the properties (\ref{10})-(\ref{12}) are valid. 

The energy budget of our present Universe includes however an uncoupled
component of non-relativistic matter at least as abundant as baryons,
namely $u^{2}\neq 0$, $u^{2}\gaq 0.05$, which measures our {}``distance''
from the solution (\ref{att}). In order to study the transient phase
of approach to the attractor we will assume that $u^{2}$ is small
enough to be compatible with a perturbative approximation, and we
linearize (\ref{sys2}) around (\ref{att}), obtaining \begin{eqnarray}
\delta x' & = & (9x_{0}^{2}-3y_{0}^{2}-4\beta x_{0}-3)\frac{\delta x}{2}-[3x_{0}y_{0}+2y_{0}(\mu +\beta )]\delta y-\beta u^{2},\nonumber \\
\delta y' & = & (\mu y_{0}+3x_{0}y_{0})\delta x+(3x_{0}^{2}-9y_{0}^{2}+2\mu x_{0}+3)\frac{\delta y}{2}.\label{lin}
\end{eqnarray}
 In this approximation, we will assume that $u^{2}$ scales in time
following the asymptotic behaviour (\ref{10}), namely \begin{equation}
u^{2}=\Om _{u}\left(a/a_{0}\right)^{3w}=\Om _{u}e^{3wN}.\label{12a}\end{equation}
 We may thus regard $u^{2}$ as a first-order {}``external source\char`\"{},
reducing (\ref{lin}) to an inhomogeneous system. The general solution
is given by \begin{eqnarray}
\delta x & = & c_{11}e^{\varepsilon _{1}N}+c_{12}e^{\varepsilon _{2}N}+A_{1}u^{2}\, ,\nonumber \\
\delta y & = & c_{21}e^{\varepsilon _{1}N}+c_{22}e^{\varepsilon _{2}N}+A_{2}u^{2}\, .\label{sol}
\end{eqnarray}

The coefficients $c_{ij}$ characterize the solution of the homogeneous
system with $u^{2}=0$. Their values depend on the initial conditions,
while the exponents $\varepsilon _{1}$ and $\varepsilon _{1}$ are
found by direct substitution to be \begin{equation}
\varepsilon _{1}=-\frac{6\beta ^{2}+9\beta \mu +3\mu ^{2}{}}{4(\beta +\mu )^{2}}-\frac{\sqrt{D}}{4(\beta +\mu )}\, ,\qquad \varepsilon _{2}=-\frac{6\beta ^{2}+9\beta \mu +3\mu ^{2}{}}{4(\beta +\mu )^{2}}+\frac{\sqrt{D}}{4(\beta +\mu )}\, .\label{eq:}\end{equation}
 The square root of \begin{equation}
D=324-32\beta ^{3}\mu -63\mu ^{2}-4\beta \mu (8\mu ^{2}-27)-4\beta ^{2}(16\mu ^{2}-45)\end{equation}
 is always imaginary in the region $2\beta >\mu $, for which acceleration
occurs. Therefore, the {}``homogeneous part'' of (\ref{sol}) describes
at the linear level the oscillatory behavior of the system while converging
to the attractor. Such oscillations clearly appear also at the level
of exact numerical solutions of the coupled equations, see for instance
\cite{4,6}. However, the amplitudes $c_{ij}$ are always subdominat
with respect to the inhomogeneous terms of Eq. (\ref{sol}), as shown
by direct numerical integration of the linear system (\ref{lin}). 

In describing the approach to the attractor the most relevant role
is thus played by the inhomogeneous contributions, proportional to
$u^{2}$. The coefficients $A_{1}$ and $A_{2}$ of Eq. (\ref{sol})
do not depend on the initial conditions, and are found to be 
\begin{eqnarray}
A_1 & = & - \frac{27 \beta}{8 \beta \mu^3 + 2(9+8 \beta^2)\mu^2 + 
4 \beta \mu (2 \beta^2 - 9) - 9 (9 +4\beta^2)}, \nonumber \\
A_2 & = & - \frac{\beta [2 \mu (\beta + \mu) -9] \sqrt{9 + 4 \beta (\beta+\mu)}}{8 \beta \mu^3 + 2(9+8 \beta^2)\mu^2 + 
4 \beta \mu (2 \beta^2 - 9) - 9 (9 +4\beta^2)} .
\end{eqnarray}
By setting $c_{ij}=0$ we can then compute, from Eq. (\ref{hprime}),
the Hubble parameter, and we obtain, to first order, \begin{equation}
2\frac{H'}{H}=-3(1+w)-6\Omega _{u}(x_{0}A_{1}-y_{0}A_{2})e^{3wN}.\end{equation}
 Direct integration leads then to the following approximate result
for the perturbed Hubble parameter: \begin{eqnarray}
H^{2}=H_{0}^{2}\, \, e^{2\int _{0}^{N}dNH'/H}= &  & H_{0}^{2}e^{-3(1+w)N}\left[1-6\Omega _{u}(x_{0}A_{1}-y_{0}A_{2})\int _{0}^{N}dNe^{3wN}\right]\nonumber \\
= &  & H_{0}^{2}\left[(1-\Omega _{R})a^{-3(1+w)}+\Omega _{R}a^{-3}\right],\label{eq:ha}
\end{eqnarray}
 where we have introduce the {}``renormalized'' fractional density
of the uncoupled component, $\Omega _{R}$, defined by \begin{eqnarray}
 &  & \Om _{R}=\Om _{u}(1+\ep ), ~~~~~~~~~
\epsilon \, \, =\, \, 6\frac{y_{0}A_{2}-x_{0}A_{1}}{3w}-1\, \, =\, \, \nonumber \\
 &  & =\frac{9(9+2\beta \mu )}{2\mu ^{2}(9+8\beta ^{2})+4\beta \mu (2\beta ^{2}-9)-9(9+4\beta ^{2})+8\beta \mu ^{3}}.\label{eq:eps}
\end{eqnarray}
Note that $\Om _{R}/\Om _{u}$ is always larger than (but near
to) unity, for realistic values of $\b {,}\mu $ \cite{4}. 
In Fig. 1 we compare the result (\ref{eq:ha}) for $H^{2}$ with a
numerical integration of the exact equations (\ref{sys2}) which includes the homogeneous part of the solution:
the errors turn out to be very small, less than 3\% in the range $z<6$. 

\begin{figure}[t]
\centerline{\includegraphics[scale=0.7,bb=0bp 520bp 598bp 743bp]{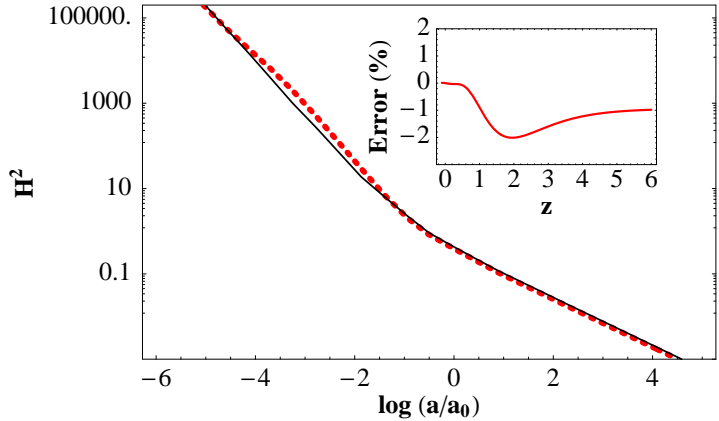}}
\caption{Approximations to $H^{2}$ (assuming $\beta =4,\mu =3,\Omega _{b}=0.05,\Omega _{c}=0.3$).
The dotted curve is the numerical result, the full curve the linear
fit. In the inset, the relative error. }
\end{figure}

The recent compilation of SNe Ia data can now be used 
to put constraints on the two parameters $\Om _{R}$ and $w$
entering the expression for $H(a)$ in Eq. (\ref{eq:ha}). Such an
expression is formally identical to that of a conventional, perfect
fluid model with a pressureless component, $\Omega _{R}$, and a dark
energy component, $1-\Omega _{R}$, with constant equation of state. We have performed a numerical analysis with the \emph{gold}
sample of data presented in \cite{9}, and we have reproduced  the same results as in \cite{9} for constant equation
of state, but we now interpret $\Omega _{R}$ as the total fraction
of uncoupled matter rescaled by the (model-dependent) factor $1+\epsilon $. 

In coupled dark energy models, characterized by Eq. (\ref{10}),  the effective equation of state of the
asymptotic regime always corresponds to a parameter $w>-1$. We thus
begin the analysis by restricting ourselves to this region of parameter
space. We also notice that, in all subsequent plots, the likelihood
function has been integrated over a constant offset of the apparent magnitude that takes into account the uncertainty on the absolute calibration both of the SN magnitude and of the Hubble constant. All results are therefore independent
of the present value of the Hubble parameter. 

In Fig. 2 we show the likelihood contours in the plane 
$(w,\Omega _{R})$. 
The value $\Omega _{R}\simeq \Om _{u}=0.05$ lies near the limit of,
but inside, the contour corresponding to the 90\% confidence level.
This shows that an uncoupled percentage of dark matter density of
the order of the baryon density is not ruled out by present SNe Ia
data. Moreover, if some fraction of dark matter is also actually uncoupled,
the fit is generally improved. However, $\Om_u$ cannot be arbitrarily close to $\Om_m$, where $\Om _{m}=\Om _{u}+\Om _{c}\simeq 0.3$ is 
the total non relativistic matter density that can be estimated through
the mass in galaxy clusters. Moving toward the higher confidence
regions of Fig. 2 requires indeed a lower and lower fraction of coupled dark
matter, but the latter must be compensated with higher and higher values
of the coupling $\beta $ in order to get an equivalent amount of
acceleration.

\begin{figure}[t]
\centerline{\includegraphics[scale=0.7, bb=0bp 400bp 512bp 702bp]{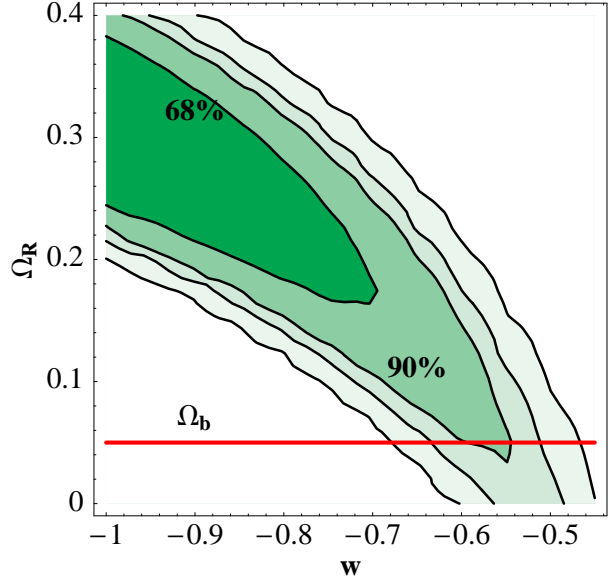}}
\caption{Contour plot obtained from the SNe Ia of \cite{9}, cutting to $w>-1$.
Likelihood contours, from inside to ouside, are at the 68\%, 90\%,
95\%, 99\% confidence level. $\Omega _{R}$ is the effective fraction
of uncoupled matter evolving as $a^{-3}$ (see Eq. (\ref{eq:ha})).
The line labelled by $\Om _{b}$ marks the present baryon fraction.}
\end{figure}

This point is better illustrated in Fig. 3, where the confidence regions are plotted
in the $\beta $--$\Omega _{u}$ plane for $\Omega _{m}=0.27$. 
For a given value of the external parameter $\Omega _{m}$, in fact,
any point of the $w-\Omega _{R}$ plane can be related to a point 
of the $\beta-\Omega _{u}$ plane,  using Eq. (\ref{eq:eps}) and the relation 
\begin{equation}
\mu (\beta ,\Omega _{m})=\frac{2\beta \Omega _{m}-\beta +\sqrt{\beta ^{2}+18(1-\Omega _{m})}}{2(1-\Omega _{m})}, 
\label{26}\end{equation}
derived
from $\Omega _{m}=1-x_{0}^{2}-y_{0}^{2}$, and  valid near the attractor  (we took the root that gives positive acceleration) .
Values of $\Omega _{m}$ different from, but near to, 0.27, lead  qualitatively to the same result. 

\begin{figure}[t]
\centerline{
\includegraphics[scale=0.7, bb=0bp 430bp 512bp 742bp]{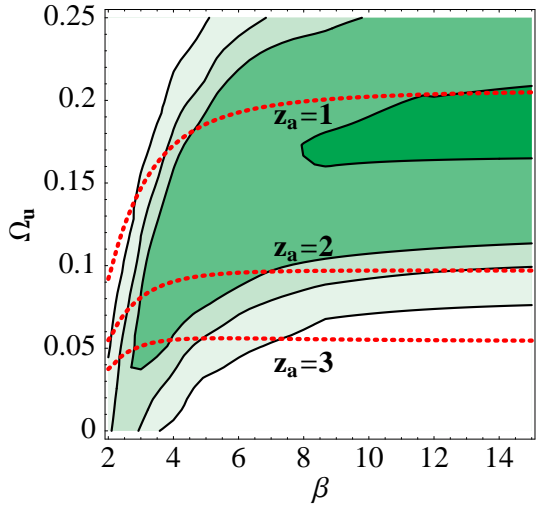}}
\caption{The SNe Ia results projected on the plane $\beta ,\Omega _{u}$,
at fixed $\Omega _{m}=0.27$. Likelihood contours, from inside to
outside, are at the 68\%, 90\%, 95\%, 99\% confidence level. The dotted
curves mark the parameter values for which $z_{\rm{acc}}$ equals
$1$, $2$ and $3$, respectively. Values of $z_{\rm{acc}}$ above
$1$ are within one $\sigma $, and even $z_{\rm{acc}}=3$ is
not excluded at more than 90\% c.l..}
\end{figure}

We can see from Fig. 3, where $\Omega _{m}$ has been fixed at a reasonable
value, that we cannot  arbitrarily shrink to zero the value of 
$\Omega _{c}$ without increasing $\beta $. A coupling $\beta $ much greater than unity, on the other hand, leads strong effective violations of the gravitational universality in the dark matter ``subsector" corresponding to $\Om_c$. Such violations are in principle constrained by present observations comparing the distribution of dark matter and baryons in galaxies and clusters \cite{grad}. The corresponding upper bounds derived on $\b$, however, depend on several assumptions concerning the dark matter distribution and are of limited
generality. Moreover, the bounds in \cite{grad} are derived assuming that all the dark
matter is coupled, while here we are taking the more general point
of view that $\Omega _{c}$ may differ from $\Omega _{m}$. 

For all these reasons, a firm quantitative limit on $\b$ is hard to be derived from present data, and will be discussed in a forthcoming paper \cite{11a}. In this paper we will not exclude \emph{a priori} large values of $\beta $, taking into account, however, that 
the large couplings required for a good fitting to SN data could impose severe constraints on coupled models of dark energy. 

 In Fig. 3 we have also plotted the curves $z_{\rm{acc}}$ marking the beginning of the accelerated regime, which are obtained from Eq. (\ref{eq:ha}) as 
\begin{equation}
z_{\rm{acc}}(\Omega _{R},w)=[(3w+1)(\Omega _{R}-1)/\Omega _{R}]^{-1/3w}-1.\label{zaccb}
\end{equation}
It follows that even $z_{\rm{acc}}=3$
cannot be excluded at more than 90\% confidence level, while a value
of $z_{\rm{acc}}>1$ is well within the 68\% confidence level.
This shows that the claim of \cite{9} (see, however, \cite{cora})
about experimental detection
of deceleration at $z>0.5$ does not apply to models of coupled dark energy. 

\begin{figure}[t]
\centerline{
\includegraphics[scale=0.7, bb=0bp 450bp 598bp 700bp]{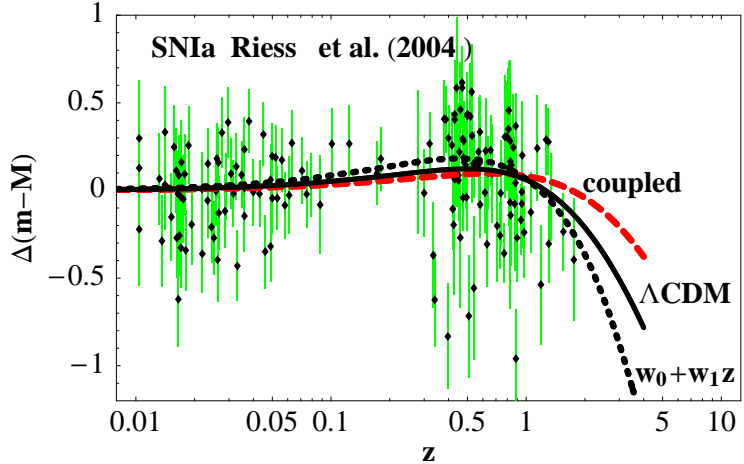}}
\caption{Best fits of the distance modulus $\Da (m-M)$ for three different
models: $\Lambda $CDM ($\Omega _{m}=0.27,\Omega _{\Lambda }=0.73$),
coupled dark energy $(w=-0.6,~ $\textbf{$\Om _{R}=0.05$} ) and a varying
equation of state $w=w_{0}+w_{1}z$, with $w_{0}=-1.31$, $w_{1}=-1.48$,
fixing $\Omega _{m}=0.27$. The data represent the \emph{gold} sample
of \cite{9}.}
\end{figure}

On a more qualitative level, we show in Fig. 4 the best fit of SNe
Ia for three classes of models, compared to the dataset of \cite{9}:
the standard $\Lambda $CDM scenario, a varying equation of state
with $w=w_{0}+w_{1}z$, and our coupled model. The plotted variable
is the distance modulus $\Delta (m-M)\equiv (m-M)_{\rm{model}}-(m-M)_{\rm{Milne}}$,
as customary (see for instance \cite{14}), where apparent and absolute
magnitude are related by 
\begin{equation}
m-M=5\log _{10}d_{L}(z), ~~~~~~~~~
d_{L}(z)=(1+z)\int _{0}^{z}dz'H^{-1}(z').\end{equation}
 As it can be seen, the coupled model is almost degenerate (difference
less than $\Delta m=0.1)$ with $\Lambda $CDM up to $z\approx 1.5$,
and with the varying-$w$ model up to $z\approx 1$. However, in these
three models, the acceleration begins at very different epochs ($z_{acc}=0.75$
for $\Lambda $CDM, $z_{acc}=0.69$ for $w=w_{0}+w_{1}z$, to be compared
to $z_{acc}=3.53$ for the coupled model). 

We can also perform the fit by including the region of parameter space
with $w<-1$. In that case, the best fit moves to larger values of
$\Omega _{u}$ (see Fig. 5). As a consequence, the baryon value $\Om _{R}\simeq \Omega _{b}\simeq 0.05$
moves near the 97\% confidence level. This shows, on one hand, that
additional SNe Ia data may have the potential to disprove models in
which all (or most of) dark matter is coupled to dark energy. On the
other hand, this also shows that if the uncoupled fraction of dark
matter is small enough, namely $\Om _{u}\laq \Om _{R}\laq 0.2$, then
we are lead to the region of parameter space with $w>-1$, excluding
{}``phantom\char`\"{} or {}``k-essence\char`\"{} models of dark
energy \cite{11}, possibly associated to embarrassing cosmic violations
of the null energy condition. 

\begin{figure}[t]
\centerline{\includegraphics[scale=0.7, bb=0bp 400bp 512bp 702bp]{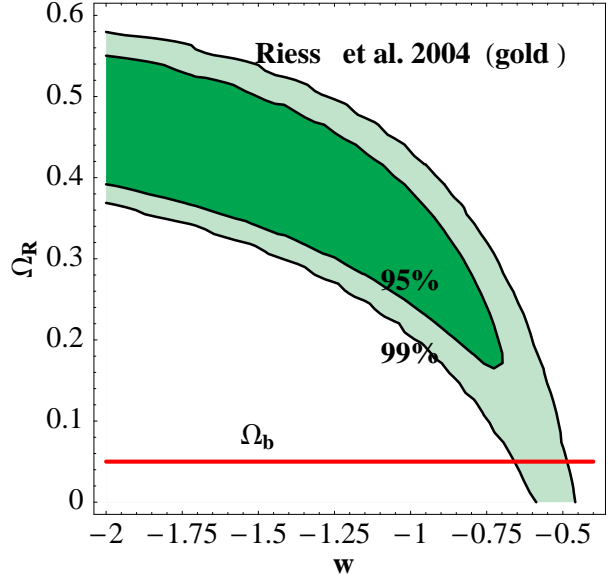}}
\caption{Contour plot obtained from the SNe Ia of \cite{9}, including $w<-1$.
Likelihood contours, from inside to ouside, are at the 95\% and 99\%
confidence level. The variables $w,\Om _{R},\Om _{b}$ are the same
as in Fig. 2. }
\end{figure}

Finally, we can also derive constraints on $w$ alone by marginalizing
over $\Omega _{R}$ (i.e. integrating the full likelihood over the
domain of $\Omega _{R}$, see Fig. 6). In the minimal case in which
the totality of (renormalized) uncoupled matter coincides with the
baryon fraction, $\Om _{b}$, we marginalize over $\Omega _{R}=0.05\pm 0.01$
(we disregard here the possible small difference between $\Omega _{u}$ and
$\Omega _{R}$). The best fit turns out to be $w=-0.6\pm 0.05$, rather
independently of the exact value of $\Omega _{R}$. If we keep instead
the range $\Omega _{R}=0.04-0.3$, i.e. the full possible range of
the uncoupled matter fraction, we then obtain $w=-0.8\pm 0.1$. This
shows that, with some finite fraction of coupled matter, the allowed
range of values of $w$ remains firmly anchored to the $w>-1$ region
of parameter space, in contrast with the standard case in which $\Omega _{R}\approx 0.3\pm 0.05$,
for which the peak of the likelihood is at $w\approx -1.1$ and the
likelihood content of the $w<-1$ region is 80\%.

\begin{figure}[t]
\centerline{\includegraphics[scale=0.6,bb=0bp 490bp 512bp 735bp]{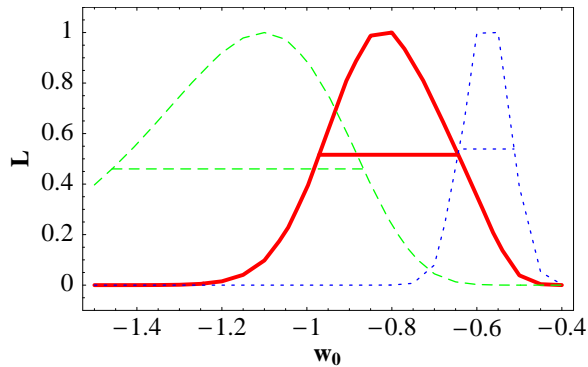}}
\caption{Likelihood for $w$ marginalized on $\Omega _{R}$, with priors in three different intervals. Full curve: $\Omega _{R}\in (0.04,0.3)$,
flat prior. Dotted curve: $\Omega _{R}\in (0.04,0.06)$, flat prior.
Dashed curve: $\Omega _{R}\in 0.3\pm 0.05$, Gaussian prior. The peak moves to $w>-1$ when $\Omega _{R}$ is smaller than 0.3. The horizontal lines mark the 68\% confidence interval.}
\end{figure}

In conclusion, we have shown that from the analysis
of recent SNe Ia there is still some room for strongly coupled
dark energy scenarios which can ease the coincidence problem, and are compatible with an earlier beginning of the accelerated regime. 
Although more conventional
models fit better the data, they seem to point toward a $w<-1$ equation
of state for the cosmic dark energy component. If this tendency is confirmed by future observations we
will face the problem of a theoretical interpretation of such a strong negative pressure. {}``Phantom'' models, in fact, seem to be plagued with
severe quantum instabilities. Coupled dark energy models, in this
respect, could provide an interesting alternative. 

We found that the {}``minimal model'' in which only the baryons
are uncoupled fits the SNe Ia at 2$\sigma $ if we restrict to $w>-1$, 
and at roughly 98\% c.l. if we include the region $w<-1$. In both
cases the best fit for the equation of state is $w=-0.6\pm 0.1$.
The universe may start the present acceleration at a very early epoch, $z\approx 3$,
much earlier than any uncoupled model. If  some fraction of
dark matter is let to be uncoupled as well, then the fit improves steadily as the fraction increases, but at the price of a large coupling $\beta $, which in turn induces potentially harmful violations of universality at the level of large scale gravitational interactions. A modest increase in the data statistics could definitely reject the minimal model of coupled dark energy, thus forcing an admixture of coupled and uncoupled dark matter as an acceptable cosmological 
scenario.

\vspace{0.5cm}
\noi
{\it Acknowledgments.} We are very grateful to  Gabriele Veneziano for many helpful suggestions, and for a fruitful collaboration during the early stages of 
this work.

\end{document}